\documentstyle[preprint,aps]{revtex}
\tightenlines

\begin{document}
\draft
\title{ Quantum entanglement of flux qubits via a resonator.}
\author{A. Yu. ~Smirnov$^{(1)}$ and A. M. ~Zagoskin$^{(1,2)}$ }

\address{$^{(1)}$D-Wave Systems, Inc., 320-1985 W. Broadway,
Vancouver, BC, Canada V6J 4Y3}

\address{$^{(2)}$ Physics and Astronomy Dept., The University of British Columbia, 6224 Agricultural Rd., Vancouver, B.C., V6T 1Z1, Canada}

\date{\today}
\maketitle

\begin{abstract}
{We show that flux qubits can be efficiently entangled by inductive coupling to a tunable resonant circuit, in the scheme reminiscent of atoms' entanglement through the optical cavity mode.  It is shown, in particular, that the single-photon excitation of the resonator produces the pure Bell state of qubits with the completely disentangled LC circuit.  }
\end{abstract}


{\bf 1.}
Successful realization of quantum computing critically depends on our
ability to entangle quantum states of qubits \cite{Nielsen1}. Controllable coupling mechanism necessary to perform this task must also allow for variation of single qubit's parameters, which is inevitable if qubits are macro- or at
least mesoscopic devices. This is the case for superconducting flux qubits
\cite{Makhlin1,Mooij1}. The alternative approach, fine tuning of single qubit
parameters, is not always applicable.

The natural way of coupling superconducting qubits is through an
intermediate resonant LC circuit (tank)\cite{Makhlin2,Yukon1,Plastina1,Blais1}. The tank circuit can be made tunable by including in it a biased Josephson junction \cite{Plastina1,Blais1}, the effective inductance of which depends on the bias current.

In the limit\cite{Makhlin2}, when the tank resonant frequency $\omega _{T}\gg
\Delta $, where $\Delta $ is a characteristic frequency of the qubit, the
role of the tank is reduced to creating a direct qubit-qubit interaction,
which can only be controlled by tuning individual qubits. Here we consider
the case when $\omega _{T}$ and $\Delta $ are comparable, so that there are
actual excitations (photons) in the tank.

The presence of actual excitations in the LC contour allows more freedom,
e.g. its {\em consecutive} coupling to the qubits, when the frequencies of
the latter differ. The problem we approach is analogous to the one
extensively studied in quantum optics \cite{Raimond1,Cirac1,Dung1,Yukon1} with respect to entanglement of atoms through the cavity modes. We concentrate here on
developing the theory of dynamic evolution of the "two qubits plus
resonator" system in the case of {\em simultaneous} coupling the qubits to the resonant LC circuit. The goal is to find the conditions which allow to
maximally entangle the qubits with each other, while disentangle them from
the quantum state of the tank. The decoherence effects are not taken into
account explicitly. We must assume anyway that the decoherence time of
qubits significantly exceeds the time necessary for manipulations with them
(which is a realistic assumption in view of recent experimental successes
with superconducting qubits\cite{Saclay1,Han1,Martinis1}). The LC circuit
itself should have the decoherence time of the same order as qubits.

{\bf 2.} We consider two phase qubits, $"a"$ and $"b"$, having the same tunneling splitting. Each of these qubits is inductively coupled to the LC circuit ("tank") tuned in resonance with the qubits. The Hamiltonian of the tank with an inductance $L_T $ and a capacitance $C_T$ has the form
\begin{equation}
H_T = {q_T^2 \over 2 C_T} + {L_T I_T^2\over 2},
\end{equation}
where $q_T $ and $I_T$ are a charge and a current in the tank, respectively. It should be noted, that for these variables we have the commutation rule: $[I_T, q_T]_- = i \hbar /L_T, $ which follows from the relation between a flux in the tank, $\Phi_T = L_T I_T, $ and the charge $q_T: [\Phi_T, q_T]_- = i\hbar.$
In a two-state approximation a persistent current in the qubit loop takes two values $\pm I_a$ (for $a-$qubit) and can be described by the Pauli matrix $\sigma_x.$ As a result, the operator of the coupling energy between the qubits and the tank will looks like:
\begin{equation}
H_{int} = - (M_a I_a \sigma_x + M_b I_b \tau_x)I_T.
\end{equation}
Here $M_a = k_a\sqrt{L_a L_T}$ stands for the mutual inductance of the tank and the $a-$qubit having the inductance $L_a$. The same is true for the b-qubit which has the inductance $L_b$ and is describes by the Pauli matrices $\tau_x, \tau_y, \tau_z.$ The qubit currents $I_a, I_b$ are constant parameters here.
The dimensionless coefficient $k$ obeys a relation: $k^2 Q \simeq 0.1 - 1 $ with $Q$ being the quality factor of the LC-circuit.

The tank represents itself a single mode cavity with excitations which we will call thereafter "photons". Because of this, it is convenient to work with creation-annihilation operators of photons in the tank $b^+, b$, where:
\begin{equation}
b = \sqrt{L_T\over 2\hbar \omega_T }(I_T + i \omega_T q_T).
\end{equation}
Then, for the electric current in the tank we have
\begin{equation}
I_T = \sqrt{\hbar \omega_T \over 2 L_T} ( b + b^+).
\end{equation}

The Hamiltonian of two qubits with the same tunneling elements $\Delta_a = \Delta$ and $\Delta_b = \Delta $ interacting with the tank can be written as
\begin{equation}
H = \Delta \sigma_z + \Delta \tau_z - (\lambda_a\sigma_x + \lambda_b \tau_x)(b + b^+) + \hbar \omega_T (b^+b + 1/2),
\end{equation}
where the coupling constants are $\lambda_a = M_a I_a\sqrt{\hbar
\omega_T /2L_T}, \lambda_b = M_b I_b\sqrt{\hbar \omega_T /2L_T}.$
Without the interaction between the qubits and the tank the
evolution of operators $\sigma_{\pm}= \sigma_x \pm i \sigma_y,
\tau_{\pm} = \tau_x \pm i\tau_y, b $ is described as
\begin{equation}
\sigma_{\pm}(t) = \exp(\pm 2 i \Delta t ) \sigma_{\pm}(0);
\tau_{\pm}(t) = \exp(\pm 2 i \Delta t ) \tau_{\pm}(0);
b(t) = e^{-i \omega_T t} b(0).
\end{equation}
To analyze the case of strong resonant coupling between the qubits and the tank, when the frequency of the tank is exactly equal to the energy splitting between the levels of the qubits, $ \omega_T = 2 \Delta,$  we use a rotating wave approximation and neglect of the fast oscillating terms in the Hamiltonian (5). As a result, we get the Jaynes-Cummings model \cite{Jaynes1} with two two-level systems  resonantly coupled to the single mode cavity. The Hamiltonian of this model has the form: $H = H_0 + H_{int},$ with
\begin{eqnarray}
H_0 = {\omega_T \over 2} \sigma_z + {\omega_T \over 2} \tau_z  + \omega_T (b^+b + 1/2),\nonumber\\
H_{int} = -{\lambda_a \over 2} (\sigma_+ b + \sigma_- b^+)  -{\lambda_b \over 2} (\tau_+ b + \tau_- b^+).
\end{eqnarray}
As in the original Jaynes-Cummings model, two parts of the Hamiltonian $H$ commute: $[H_0, H_{int}] = 0.$ It means that the operators $H_0$ and $H_{int} $ have the common system of eigenfunctions.
Let's introduce $\alpha_a, \beta_a $ as eigenfunction of the $\sigma_z-$ matrix:$\sigma_z \alpha_a = \alpha_a, \sigma_z \beta_a = - \beta_a;$ and $ \alpha_b, \beta_b $ as eigenfunctions of the z-matrix of the second qubit, $\tau_z,$ with  eigenvalues $\pm 1.$ Then, the action of operators $\sigma_{\pm}$ on these functions is described as follows: $\sigma_+ \alpha_a = 0, \sigma_+ \beta_a = 2 \alpha_a; \sigma_- \alpha_a = 2 \beta_a, \sigma_- \beta_a = 0.$ The similar is true for the second qubit. The states of the LC circuit are represented by the eigenfunctions of the harmonic oscillator, $|n\rangle ,$ corresponding to the eigen-energy $\omega_T(n+1/2).$ The action of the creation-annihilation operators on these states is well-known: $b|n\rangle = \sqrt{n} |n-1\rangle, b^+|n\rangle = \sqrt{n+1}|n+1\rangle .$
We choose the following eigenstates of the Hamiltonian $H_0$ (7) corresponding to the same eigenenergy $n\omega_T, H_0 \Psi_j = n\omega_T \Psi_j, j = 1,..4$:
\begin{eqnarray}
\Psi_1 = |n-1\rangle \alpha_a \alpha_b, \Psi_4 = |n+1\rangle \beta_a \beta_b, \nonumber\\
\Psi_2 = |n\rangle \alpha_a \beta_b, \Psi_3 = |n\rangle \beta_a \alpha_b.
\end{eqnarray}
Here the number $n$ is counted from 1. In the case when $n=0$ the wave function of the first state is assumed to be zero, $\Psi_1 = 0,$ and we have only three basis states. If $n=1$, then in the first state we have no photons, but both qubits are in their upper states contributing $(0.5 + 0.5)\omega_T = \omega_T$ to the total energy (we do not consider here the tank-qubits interaction). In the second state,$ \Psi_2,$ the second qubit turns over, from the state $\alpha_b$ with the energy $\omega_T /2$ to the state $\beta_b$ with the energy $-\omega_T /2.$
The released energy goes into the tank creating one photon having the energy $\omega_T.$ The same process occurs in going from state 1 to state 3 when the first qubit turns over. Due to the commutativity of $H_0$ and $H_{int}$ the action of the coupling Hamiltonian $H_{int}$ on the eigenfunctions (8) of the free Hamiltonian $H_0$  does not fall outside the basis  $\Psi_1, ..\Psi_4:$
\begin{eqnarray}
H_{int}\Psi_1 = - \sqrt{n} (\lambda_a \Psi_3 + \lambda_b \Psi_2 ), \nonumber\\
H_{int}\Psi_2 = -  (\lambda_a \sqrt{n+1} \Psi_4 + \lambda_b \sqrt{n} \Psi_1 ), \nonumber\\
H_{int}\Psi_3 = -  (\lambda_a \sqrt{n} \Psi_1 + \lambda_b \sqrt{n+1} \Psi_4 ), \nonumber\\
H_{int}\Psi_4 = - \sqrt{n+1} (\lambda_a \Psi_2 + \lambda_b \Psi_3 ),
\end{eqnarray}
or, that is the same, the nonzero matrix elements of the operator $H_{int}$ in the basis $\Psi_1, ..\Psi_4$ will look like:
\begin{eqnarray}
\langle 1|H_{int}|2\rangle  = \langle 2|H_{int}|1\rangle = - \sqrt{n} \lambda_b;
\langle 1|H_{int}|3\rangle = \langle 3|H_{int}|1\rangle = - \sqrt{n} \lambda_a; \nonumber\\
\langle 2|H_{int}|4\rangle = \langle 4|H_{int}|2\rangle = - \sqrt{n+1} \lambda_a;
\langle 3|H_{int}|4\rangle = \langle 4|H_{int}|3\rangle = - \sqrt{n+1} \lambda_b.
\end{eqnarray}
The eigenfunctions $ \{\Psi \}$ of the total Hamiltonian $H = H_0 + H_{int}, H\Psi = E \Psi ,$ can be  represented as a superposition of the basis function $\Psi_1, ..\Psi_4: \Psi = \sum_{j=1}^{j=4} c_j \Psi_j.$ As a result we obtain the system of four linear equations for the coefficients $c_1,.. c_4.$ After some transformations these equations take the form:
\begin{eqnarray}
(E - n \omega_T ) c_1 = - \sqrt{n} (\lambda_b c_2 + \lambda_a c_3), \nonumber\\
(E - n \omega_T ) c_4 = - \sqrt{n+1} (\lambda_a c_2 + \lambda_b c_3),
\end{eqnarray}

\begin{eqnarray}
\{(E - n \omega_T )^2 - \lambda_b^2 n - \lambda_a^2 (n+1)\}c_2 - \lambda_a \lambda_b (2n + 1) c_3 = 0, \nonumber\\
\{(E - n \omega_T )^2 - \lambda_a^2 n - \lambda_b^2 (n+1)\}c_3 - \lambda_a \lambda_b (2n + 1) c_2 = 0.
\end{eqnarray}
This system has nontrivial solutions when the the corresponding determinant is equal to zero:
\begin{equation}
Det = (E - n\omega_T)^4 -(\lambda_a^2 + \lambda_b^2)(2n+1)(E - n\omega_T)^2 +
(\lambda_a^2 - \lambda_b^2)^2 n(n+1) = 0.
\end{equation}
This condition gives us a set of four eigen-energies of the total Hamiltonian $H$
\begin{eqnarray}
E^{(1,2)} = n \omega_T \pm\sqrt{(\lambda_a^2 + \lambda_b^2)(n +1/2) +
\sqrt{(\lambda_a^2 + \lambda_b^2)^2(n +1/2)^2 - (\lambda_a^2 - \lambda_b^2)^2
n(n + 1)}}, \nonumber\\
E^{(3,4)} = n \omega_T \pm\sqrt{(\lambda_a^2 + \lambda_b^2)(n +1/2) -
\sqrt{(\lambda_a^2 + \lambda_b^2)^2(n +1/2)^2 - (\lambda_a^2 - \lambda_b^2)^2
n(n + 1)}}.
\end{eqnarray}

{\bf 3.} Let's proceed to a consideration of some particular examples.
In the case when the interaction of the tank with the second qubit is absent $(\lambda_b = 0)$ for the energy levels $E^{(\nu)}$  and for the corresponding eigenstates $\Psi^{(\nu )}, \nu = 1,..4,$  we obtain, respectively:
\begin{eqnarray}
E^{(1,2)} = n \omega_T \pm \lambda_a\sqrt{n+1};
E^{(3,4)} = n \omega_T \pm \lambda_a\sqrt{n}. \nonumber\\
\Psi^{(1,2)} = {1\over \sqrt{2}} \left(|n\rangle \alpha_a \mp |n+1\rangle \beta_a \right) \beta_b,\nonumber\\
\Psi^{(3,4)} = {1\over \sqrt{2}} \left(\mp |n+1\rangle \alpha_a + |n\rangle \beta_a \right) \alpha_b.
\end{eqnarray}
It is evident, that {\em the second qubit is completely disentangled with the tank as well as with the first qubit}. Nevertheless, {\em the first qubit and the tank are strongly entangled} in this situation.

It can be expected that the maximal entanglement could be reached when both qubits are coupled to the
tank with the same strength, $\lambda_a = \lambda_b = \lambda = k \sqrt{\omega_T L_q I_q^2/2},$ where
$L_q $ is a qubit inductance, and $I_q$ is a current in a qubit's loop. Then the first and the second
energy levels are different,
\begin{equation}
E^{(1,2)} = n \omega_T \pm 2 \lambda \sqrt{n+1/2},
\end{equation}
whereas the third and the fourth states have the same energies
\begin{equation}
E^{(3,4)} = n \omega_T.
\end{equation}
The corresponding orthogonal wave functions take the form
\begin{eqnarray}
\Psi^{(1,2)} = {1\over 2}\left[ \sqrt{n \over n + 1/2}|n - 1\rangle \alpha_a \alpha_b +
\sqrt{n + 1 \over n + 1/2}|n + 1\rangle \beta_a \beta_b \mp |n\rangle  \alpha_a \beta_b \mp |n\rangle \beta_a \alpha_b \right], \nonumber\\
\Psi^{(3,4)} = {1\over 2}\left[ \sqrt{n + 1 \over n + 1/2}|n - 1\rangle \alpha_a \alpha_b -
\sqrt{n \over n + 1/2}|n + 1\rangle\beta_a \beta_b \pm |n\rangle \alpha_a \beta_b \mp |n\rangle \beta_a \alpha_b \right].
\end{eqnarray}
Apparently, now we have strongly entangled qubits and the tank. However, { \em if the number of photons in the LC-circuit is very high}, $ n\gg 1,$ so that $|n\pm 1 \rangle \simeq |n \rangle, $ the system is actually in a factorized state :
\begin{equation}
\Psi^{(1,2)} = {1\over 2}|n \rangle (\alpha_a \mp \beta_a )(\alpha_b \mp \beta_b ),
\Psi^{(3,4)} = {1\over 2}|n \rangle (\alpha_a \mp \beta_a )(\alpha_b \pm \beta_b ),
\end{equation}
as expected, since there should be {\em no entanglement through a classical
system}.

{\bf 4.} It is of interest to trace the evolution of the originally non-entangled state, say, $\Psi(0) = \Psi_0 = |0\rangle \alpha_a \alpha_b.$ In this initial state both qubits are in the up-state and there are no photons inside the tank.
The time evolution of the wave function $\Psi(t)$ in the Schr\"{o}dinger picture is described by the superposition:
\begin{equation}
\Psi(t) = \sum_{\nu =1}^{\nu =4}\sum_{n=1}^{\infty} u_{\nu } \exp(-i E^{(\nu )}t)\Psi^{(\nu)},
\end{equation}
where $E^{(\nu)},\Psi^{(\nu)}$ are the eigen-energies (16), (17) and eigenstates (18) of our system. These states and energies depend on the number of photons in the tank $n$. Because of this we have a sum over $n$ in Eq.(20).
The coefficients $u_{\nu}$ in the superposition (20) are determined by the relation
\begin{equation}
u_{\nu } = \langle \Psi^{(\nu )} |\Psi (0)\rangle ,
\end{equation}
and also depend on the number of photons. For our choice of the initial state the quantum number it should be: $n=1, $ and $u_1=u_2 = 1/\sqrt{6}, u_3=u_4 = 1/\sqrt{3}.$  As a result the following four states participate in the evolution of our wave function $\Psi(t):$
\begin{eqnarray}
\Psi^{(1)} = {1 \over \sqrt{6}}|0\rangle \alpha_a \alpha_b +
{1 \over \sqrt{3}}|2\rangle \beta_a \beta_b -
{1 \over 2 }|1\rangle (\alpha_a \beta_b + \beta_a \alpha_b), \nonumber\\
\Psi^{(2)} = {1 \over \sqrt{6}}|0\rangle \alpha_a \alpha_b +
{1 \over \sqrt{3}}|2\rangle \beta_a \beta_b +
{1 \over 2 }|1\rangle (\alpha_a \beta_b + \beta_a \alpha_b), \nonumber\\
\Psi^{(3)} = {1 \over \sqrt{3}}|0\rangle \alpha_a \alpha_b -
{1 \over \sqrt{6}}|2\rangle \beta_a \beta_b +
{1 \over 2 }|1\rangle (\alpha_a \beta_b - \beta_a \alpha_b), \nonumber\\
\Psi^{(4)} = {1 \over \sqrt{3}}|0\rangle \alpha_a \alpha_b -
{1 \over \sqrt{6}}|2\rangle \beta_a \beta_b -
{1 \over 2 }|1\rangle (\alpha_a \beta_b - \beta_a \alpha_b),
\end{eqnarray}
with the energies counted from the vacuum energy $\omega_T /2$:
$E^{(1)} = \omega_T + \lambda\sqrt{6}  , E^{(2)} = \omega_T -
\lambda\sqrt{6}, E^{(3)} = E^{(4)} = \omega_T ,$ respectively.
After the time interval $t$ our system "two qubits + tank" will be
in the state
\begin{eqnarray}
\Psi (t) = e^{-i\omega_T t} {2 + \cos(\sqrt{6}\lambda t)\over 3} |0\rangle \alpha_a \alpha_b -
e^{-i\omega_T t}\sqrt{2}{1 - \cos(\sqrt{6}\lambda t)\over 3}
|2\rangle \beta_a \beta_b - \nonumber\\
{i \over \sqrt{6}}  e^{-i\omega_T t}\sin (\sqrt{6}\lambda t)|1\rangle (\alpha_a \beta_b + \beta_a
\alpha_b).
\end{eqnarray}
It is clear from this formula that amplitudes to find the system in one or another state oscillate in time with the frequency
\begin{equation}
\Omega_R = \lambda\sqrt{6} =  k \sqrt{3\omega_T {L_q I_q^2}},
\end{equation}
where  $I_q$ is the current in the qubit's loop, and $k$ is the coupling coefficient between the qubit loop (with an inductance $L_q$) and the tank (with an inductance $L_T$), so that the mutual inductance one of the qubits and the tank, $M,$ is: $M= k \sqrt{L_q L_T}.$  The initial state, $\Psi_0 = |0\rangle \alpha_a \alpha_b$ revives in the moments $t= 2\pi m /\Omega_R, m = 0,1,2,..$
In the moments $t= (\pi/2 + \pi m) /\Omega_R$ all states of the qubits are mixed:
\begin{equation}
\Psi_{\pi/2} (t) = e^{-i\omega_T t}\left[ {2\over 3} |0\rangle \alpha_a \alpha_b  -{\sqrt{2}\over 3}|2\rangle \beta_a \beta_b  - (-1)^m{i \over \sqrt{3}}|1\rangle {\alpha_a \beta_b + \beta_a \alpha_b\over \sqrt{2}}\right].
\end{equation}
After a 1/2-period shift from the origin (at $ t= (\pi + 2\pi m) /\Omega_R)$ we have an entanglement of two up-states of the qubits, $\alpha_a\alpha_b$, two down-states, $\beta_a \beta_b$, with the vacuum state  $|0\rangle $   and with the two-photon state $|2\rangle $  of the tank:
\begin{equation}
\Psi_{\pi }(t) = e^{-i\omega_T t}\left[ {1\over 3} |0\rangle \alpha_a \alpha_b
- {2\sqrt{2}\over 3}|2\rangle \beta_a \beta_b \right].
\end{equation}

It should be noted that in the states (25), (26)  the qubits are entangled not only to one another but to the tank as well.
The same is true when we choose a two-photon initial state $\Psi ( 0 ) = |2\rangle \beta_a \beta_b.$ Now we have $u_1 = u_2 = 1/\sqrt{3}, u_3 = u_4 = - 1/\sqrt{6},$ and
\begin{eqnarray}
\Psi (t) = -e^{-i\omega_T t}\sqrt{2} {1 - \cos(\sqrt{6}\lambda t)\over 3} |0\rangle \alpha_a \alpha_b +
e^{-i\omega_T t}{1 + 2\cos(\sqrt{6}\lambda t)\over 3} |2\rangle \beta_a \beta_b + \nonumber\\
{i \over \sqrt{3}}  e^{-i\omega_T t}\sin (\sqrt{6}\lambda t)|1\rangle (\alpha_a \beta_b + \beta_a
\alpha_b).
\end{eqnarray}
Here again we are not able to get rid of the entanglement of qubits with the tank.

{\bf 5.} With this goal in mind we consider the case when our system "two qubits + tank" has an energy equal to the energy of the single photon and can be in a superposition of three states $(n=0):$
\begin{equation}
\Psi_2 = |0\rangle \alpha_a \beta_b, \Psi_3 = |0\rangle \beta_a \alpha_b,
\Psi_4 = |1\rangle \beta_a \beta_b.
\end{equation}
Now it is not enough the energy to excite the first state $\Psi_1 $ (8).
The total Hamiltonian $H = H_0 + H_{int} $ (7) has three eigen-energies (we drop the vacuum energy $\omega_T/2$)
\begin{equation}
E^{(1)} = \sqrt{2} \lambda, E^{(2)} = -\sqrt{2} \lambda, E^{(3)} = 0,
\end{equation}
with the corresponding wave functions
\begin{eqnarray}
\Psi^{(1)} =  |0\rangle {\alpha_a \beta_b + \beta_a \alpha_b \over 2 } -
{1 \over \sqrt{2}} |1 \rangle \beta_a \beta_b ,\nonumber\\
\Psi^{(2)} =  |0\rangle {\alpha_a \beta_b + \beta_a \alpha_b \over 2 } +
{1 \over \sqrt{2}} |1 \rangle \beta_a \beta_b ,\nonumber\\
\Psi^{(3)} =  |0\rangle {\alpha_a \beta_b - \beta_a \alpha_b \over \sqrt{2} }.
\end{eqnarray}

For the non-symmetric initial state, $ \Psi (0) = |0\rangle \alpha_a \beta_b ,$  when there are no photons in the tank and  the first qubit is in the excited state,  the evolution  of the wave function of the system is described by the expression
\begin{equation}
\Psi (t) = {1 + \cos (\sqrt{2} \lambda t)\over 2 } |0 \rangle \alpha_a \beta_b -
{1 - \cos (\sqrt{2} \lambda t)\over 2 } |0 \rangle \beta_a \alpha_b +
{i\over \sqrt{2}} \sin(\sqrt{2} \lambda t)|1\rangle  \beta_a \beta_b.
\end{equation}
Again, we are not able to obtain the entanglement of the qubits without the intrusion of the tank states.

However, for the symmetric excitation,
\begin{equation}
\Psi (0) = |1\rangle  \beta_a \beta_b,
\end{equation}
when we have a single photon in the tank and both qubits are in the down states, we obtain $u_1 = -1/\sqrt{2},
u_2 = 1/\sqrt{2}, u_3 = 0, $ so that the wave function of the system in the moment $t$ will look like
\begin{equation}
\Psi (t) = \cos (\sqrt{2} \lambda t) |1 \rangle \beta_a \beta_b +
i \sin(\sqrt{2} \lambda t) |0\rangle {\alpha_a \beta_b + \beta_a \alpha_b \over \sqrt{2}}.
\end{equation}
In the moments of time
$t= (\pi/2 + \pi m) /\omega_R$ {\em the single photon initially located in the LC-circuit produces
the pure Bell state of two qubits},
\begin{equation}
\Psi_{ab} = i(-1)^m |0\rangle {\alpha_a \beta_b + \beta_a \alpha_b\over \sqrt{2}},
\end{equation}
{\em with the completely disentangled tank}. Here $m = 0,1,2,..$ and the frequency of Rabi oscillations is:
\begin{equation}
\omega_R = \sqrt{2}\lambda = k \sqrt{\omega_T {L_q I_q^2}}.
\end{equation}
This process resembles the two-slit interference of the single-photon excitation which propagates into two qubits (slits) simultaneously.

{\bf 6.} In the case of Mooij type qubit \cite{Mooij1} when the persistent current in the qubit loop $I_q \simeq $ 450 $nA$,
a coupling coefficient $k \sim 1/30,$
a qubit loop inductance $L_q \sim 25 pH,L_q I_q^2 \simeq 6 \times 10^{-24} J,$
and the resonant frequency of LC-circuit $f_T = \omega_T/2\pi = 10^9 $ Hz, or
$\hbar \omega_T = 6.3 \times 10^{-25} J,$
we have $\lambda \simeq 0.5 \times 10^{-25} J, \hbar\omega_R = 1.4 \times 10^{-25} J,$ and the Rabi frequency $ f_R = \omega_R /2\pi = 1.1 \times 10^8 Hz \simeq 0.1 f_T.$ Thus, the pure Bell state is formed for the time interval
$$t_{Bell} = {1 \over 4 f_R} = 2.3 \times 10^{-9} sec.$$
This operation time is much less than the excitation lifetime, $\tau_{ph}$, in the LC-circuit with the quality factor $Q = 1000: \tau_{ph} = 10^{-6} sec,$ as well as less than the estimated decoherence time of the phase qubits \cite{Mooij1}.

In conclusion, in the framework of Jaynes-Cummings model we have investigated the protocol for entanglement of phase qubits inductively coupled to a resonant tank. It is shown that in the case of simultaneous coupling the single-photon excitation of the tank creates the maximally entangled Bell state of qubits with the completely disentangled vacuum state of the resonator. The results are applicable to the case of capacitive coupling, as well as to the qubit entanglement through optical cavities.

\bigskip

\acknowledgements{ We are grateful to M.H.S. Amin, A. Blais and A.
Maassen van den Brink for many inspiring discussions and helpful
comments on the manuscript.}

\end{document}